# Multispectral UV Imaging on Capacitive CMOS Arrays Enabled by Solution-Processed Metal-Oxide Nanoparticles

*Suman Kundu[1,2]\*, Tao Shen[1,2], Kai Betlem[1,2], Murali K Ghatkesar[2], Peter G Steeneken[2], Frans P Widdershoven[1,3]\**

1. Department of Microelectronics, Faculty of Electrical Engineering, Mathematics and Computer Science, TU Delft, Netherlands
2. Department of Precision and Microsystems Engineering, Faculty of Mechanical Engineering, TU Delft, Netherlands
3. NXP Semiconductors, Technology & Operations/CTO office, High Tech Campus 46, Eindhoven, Netherlands

## Abstract

Ultraviolet (UV) imagers are important for a variety of applications, such as quality inspection in the semiconductor industry, forensics and food quality inspection, but are often costly because they require dedicated semiconductor process flows. Here, an imaging chip is introduced that has been fabricated using standard 40 nm complementary metal-oxide-semiconductor (CMOS) technology. Instead of using a conventional charge-based photodetection principle, the imager uses a capacitive operation principle where UV-light causes capacitance changes via the photodielectric effect in a functionalization layer, which are measured by the underlying CMOS circuitry. This spin-coated or inkjet-printed functionalization layer consists of solution-processed, wide-bandgap, semiconducting metal-oxide nanoparticles, and facilitates multispectral imaging. The sensors exhibit low noise-equivalent powers (17-138 fW $Hz^{-1/2}$) across the UV bands. Unlike conventional silicon CMOS imagers, the present capacitive-CMOS platform is inherently visible-blind, providing selective UV detection. This work positions late-functionalized capacitive-CMOS arrays as a route toward reducing the cost of UV imagers, which can lead to their more widespread implementation in consumer and low-volume application-specific products.

## 1. Introduction

CMOS image sensors form the backbone of modern imaging technologies, providing low cost, high scalability, and seamless integration with on-chip electronics,[1-4] and have revolutionized consumer, industrial, automotive and biomedical applications. Next-generation consumer electronics demand a shift from conventional imaging to enhanced vision-perception, requiring sensors made with uncomplicated fabrication technologies, broad spectral sensitivity, and multispectral data integration.[5-14] However, the detection capability of current CMOS image sensors is intrinsically limited by the operation range of the underlying silicon photodiodes, which sharply decreases in the ultraviolet (UV) and infrared (IR) wavelengths owing to poor quantum efficiency.[1-5,8,10,12] Although CMOS-compatible photodetectors have made significant progress in the visible to short-wave IR regions,[8,9,11,15] their performance in the UV domain still lags behind.[10] The accurate and selective detection of UV radiation is of great importance, with applications spanning many domains, from quality inspection to food quality monitoring, astronomical and biological imaging, forensics analysis and currency validation.[10,16-20] Several approaches have been explored to extend CMOS photodetectors into the UV and to achieve spectral selectivity. A common practice is to use narrowband UV filters to block unwanted wavelengths, however, they suffer from lowering of signal intensity, limited band selectivity and added complexity for integration with CMOS sensors.[21] Recent approaches have focused on methods such as the use of downconverters[12,22,23] or the integration of wide bandgap materials in Si-CMOS image sensors.[24-33] Downconverters can enable UV detection by converting UV photons to visible light, but suffer from low conversion efficiency, reduced spatial resolution, broad emission spectra and low-stability. Wide bandgap materials such as ZnO,[24] SiC,[25] GaN,[26,27] AlGaN,[28-31] $Ga_2O_3$,[32] and thiazolothiazole-based small molecules[33] offer decent UV sensitivity when integrated with CMOS circuitry, but many of these materials are incompatible with standard CMOS processes, respond broadly across the visible and UV spectra, and require dedicated cleanroom fabrication facility which increase cost and complexity. For example, $Al_xGa_{1-x}N$-based solar-blind UV focal plane arrays (FPA) are typically realized by hybrid integration of an AlGaN detector array with a CMOS readout integrated circuit.[28-31] However, this configuration requires complex back-end integration schemes and is constrained by the limited thermal budget of CMOS processing, making monolithic integration and large-scale fabrication challenging. Due to these drawbacks, existing CMOS-based UV imagers have thus far fallen short of providing true band-selective detection within a single platform. This gap highlights the urgent need for compact, robust, low-cost and fully CMOS-compatible solutions for high-performance multispectral UV imaging. Moreover, since spectral requirements are strongly use-case dependent, a technology that is easily customizable to specific applications by postprocessing is preferred.

Here we introduce a versatile approach for multispectral UV detection and band-selective UV imaging using CMOS pixelated capacitive sensor (PCS) arrays.[34-39] PCS technology has previously been used for cellular signal detection, microparticle sensing, and oil-water microinterfacial characterization in the 1–100 MHz range.[34-39] We demonstrate that UV sensitivity is achieved by depositing wide-bandgap metal-oxide (MOX) nanoparticle (NP) dispersions onto capacitive sensor pixels, enabling selective detection of UV-A, UV-B, and UV-C bands on a single prefabricated CMOS chip. The MOX-functionalized sensors exhibit readily detectable capacitance-increments even under low-intensity UV illumination (0.5 µW cm$^{-2}$). We employ solution-processed inkjet printing and spin coating techniques to produce multispectral UV detectors and uniform imagers and thereby eliminate the need for high-temperature, sophisticated material-deposition and postprocessing steps. The devices exhibit low noise-equivalent power (NEP, 17-138 fW Hz$^{-1/2}$, Table T1, Supporting Information) across the UV bands, and unlike conventional silicon CMOS imagers, our capacitive-CMOS platform is inherently visible-blind. These fabrication strategies highlight the potential of the PCS platform to realize compact, robust, low-cost multispectral UV imagers, bringing their costs to a level where they might be incorporated in smartphones, allowing everyday users to inspect e.g., surface cleanliness, food molds, fruit quality, banknote authenticity, sunscreen and makeup coverage.

## 2. Results and Discussion

### 2.1. Capacitive-CMOS UV image sensor operation

To fabricate the multispectral UV detectors, we use PCS arrays (**Figure 1**). The monolithic 40 nm CMOS chip contains three separate sensing matrices, each comprising 1024 sensor pixels arranged in a 32×32 layout on a pseudo-hexagonal grid (Figure 1a,b, Supplementary Figures S1, S2, Supporting Information). Matrices 1 and 2 feature square electrodes with dimensions 5×5 µm$^2$ and nearest neighbor pitches of 15 and 10 µm, respectively, while Matrix 3 comprises electrodes with dimensions 4×4 µm$^2$ and a 10 µm pitch (Figure S2, Supporting Information). Figure 1b demonstrates an isometric view optical micrograph of a selected region of Matrix 1, exhibiting a 3D-like perspective of the pillar-shaped electrodes which form the PCS array and the trenches between the electrodes separating them. The pillar-shaped electrodes are designed to increase the available volume of functional materials for depositing inside trenches, which have a depth of ∼2.25 µm (Figure S3, Supporting Information). To validate the construction of a single electrode, we obtained a cross-sectional scanning electron microscopy image of the electrode via targeted focused ion beam milling (Figure 1c). The electrode structure comprises an aluminum core that acts as the conductive electrode, over which first $SiO_2$ (~330 nm) and thereafter $Si_3N_4$ (~365 nm) layers are deposited. These dielectric layers serve dual purposes; as chemical protection against corrosive environments and electrical insulation control to define the effective sensing capacitance. The sense electrodes are aligned directly above the CMOS switching transistor pairs, which are linked through ultra-short vertical stacks made from minimal-area intermediate metal islands and vias to minimize parasitic capacitances. A pair of NMOS/PMOS transistor switches are used to connect each electrode (Figure 1d) to a square wave modulation signal, and to connect surrounding electrodes to ground. The average charging current per modulation signal cycle is measured and translated into an equivalent electrode capacitance. A comprehensive explanation of the PCS chip's working principle is provided in Supplementary Note 1 (Figures S4, S5, S6, Supporting Information).

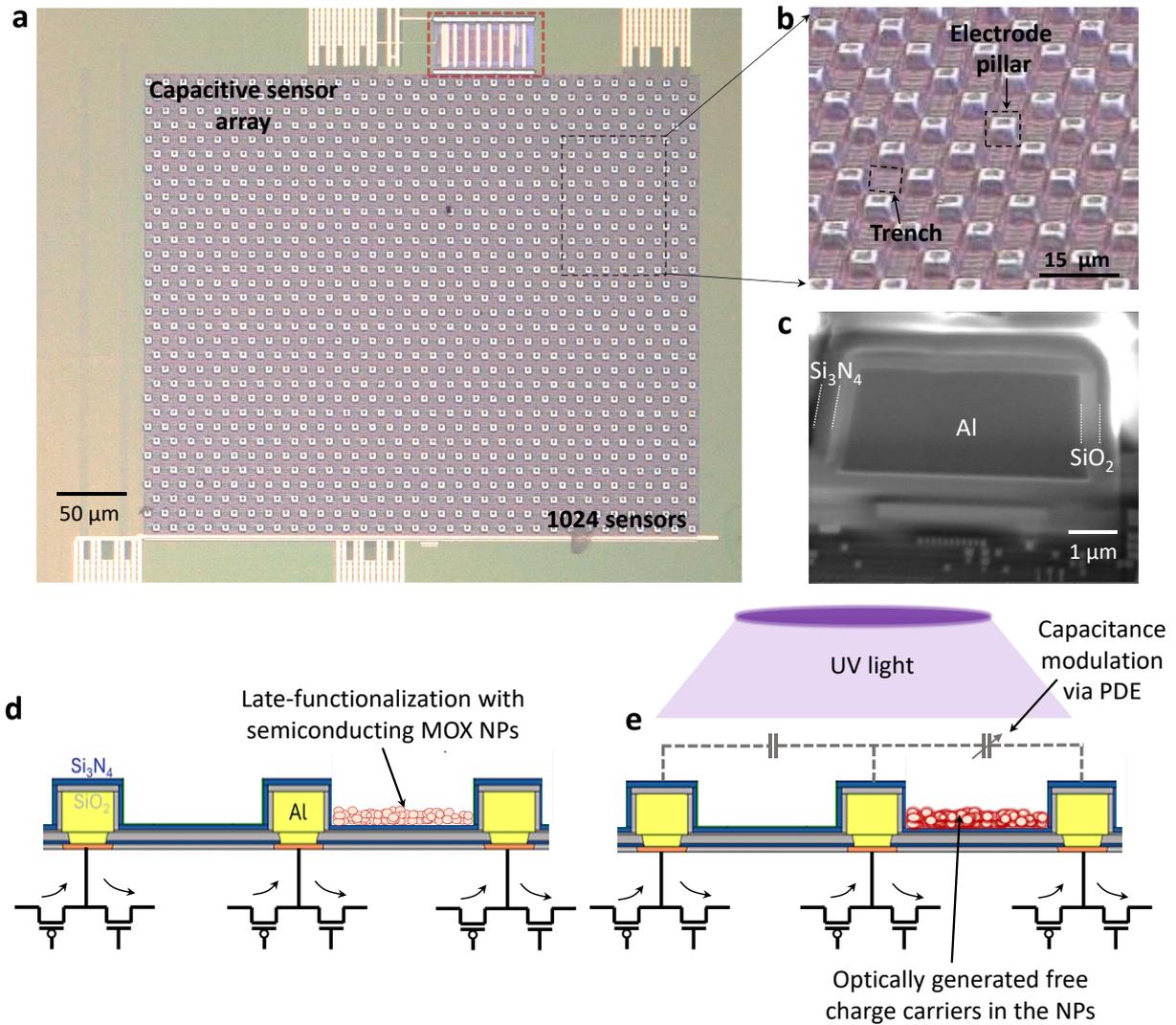

**Figure 1.** Overview of the capacitive sensor array and its operation principle. a) Optical microscopy image of the CMOS pixelated capacitive sensor (PCS) chip showing Matrix 1 which contains 1024 individually addressable pixels. The regions of read-out circuit (consisting P- and N-current mirrors) are marked with red-dashed rectangles. b) Isometric view of the sensor array under an optical microscope. The structures of pillar-like electrodes and trenches between the electrodes are indicated inside. c) Cross-sectional scanning electron microscopy image of an electrode-pillar after targeted focused ion beam milling. The electrode consists of an Al metal island, covered by a passivation layer stack of $SiO_2$ and $Si_3N_4$. d) Cross-sectional schematic of the individual sensor pixels. To measure an electrode's capacitance, on-chip PMOS/NMOS switch transistors are used to connect it to a square wave modulation signal and to connect the surrounding electrodes to the ground. The average charging current per modulation signal cycle is measured and, with known modulation voltage amplitude, translated into an equivalent electrode capacitance. A section of the prefabricated chip can be functionalized with high-bandgap semiconducting metal-oxide nanoparticles (MOX NPs). e) Exposure to UV light with a photon energy exceeding the bandgap of the MOX NPs generates free charge carriers that change the dielectric properties of the NPs and, consequently, the electrode capacitance via photodielectric effect (PDE).

To fabricate multispectral UV detectors, we deposit high bandgap semiconducting MOX NPs inside the trenches (see later sections). Figure 1d shows a part of the PCS array that is functionalized by MOX NPs. The NPs fill the interpillar trenches, creating an optically active layer inside the sensing region. Illumination by UV light excites free charge carriers in the MOX NPs (Figure 1e), increasing their conductivity. These carriers alter electric fields in the functionalization layer, and consequently modulate the capacitance between the sensing electrodes. This change in capacitance under light exposure in semiconductors, commonly referred to as the photodielectric effect (PDE), was first reported by Lenard and Saeland[40] and has been observed in various materials.[41-44] PDE can arise from either the excitation of electrons or holes across the bandgap, or the excitation of charges trapped in the bandgap.[41-44] Alternatively, in strongly light-absorbing semiconductors, the Maxwell-Wagner effect[45] can dominate, where light-induced conductivity near the material surface reduces the effective insulating thickness, thereby increasing the capacitance. Here, we leverage the PDE in the NP filled dielectric functionalization layer, to convert a CMOS capacitive sensor array into a UV image sensor.

## 2.2. Multispectral UV detection by nanoparticle functionalization

To functionalize the capacitive CMOS sensor for UV detection, we prepared aqueous dispersions of semiconducting MOX NPs - ZnO, $SnO_2$, and $Ga_2O_3$; chosen for their bandgap energies spanning the UV spectrum (Figure S7, Supporting Information). This selection allowed the platform to cover the full range of UV-A, UV-B, and UV-C wavelengths (**Figure 2a**). To characterize the intrinsic behavior of each material, we first fabricated dedicated chips coated exclusively with one oxide, ensuring that responsivity and noise could be assessed without cross-material interference (Figures S8, S9, S10, Supporting Information). For direct comparison, we also drop-coated each dispersion onto separate regions of the same PCS array (Supplementary Note 1, Figure S11, Supporting Information). The photodetection measurements reveal the spectral selectivity of the MOX films: ZnO exhibited broad UV sensitivity with a peak response under UV-A, $SnO_2$ responded mainly to wavelengths below UV-B, and $Ga_2O_3$ was active exclusively under UV-C (Figure 2b, Figure S12, Supporting Information). Notably, all devices remained insensitive to visible light (Figure S12a,b, Supporting Information).

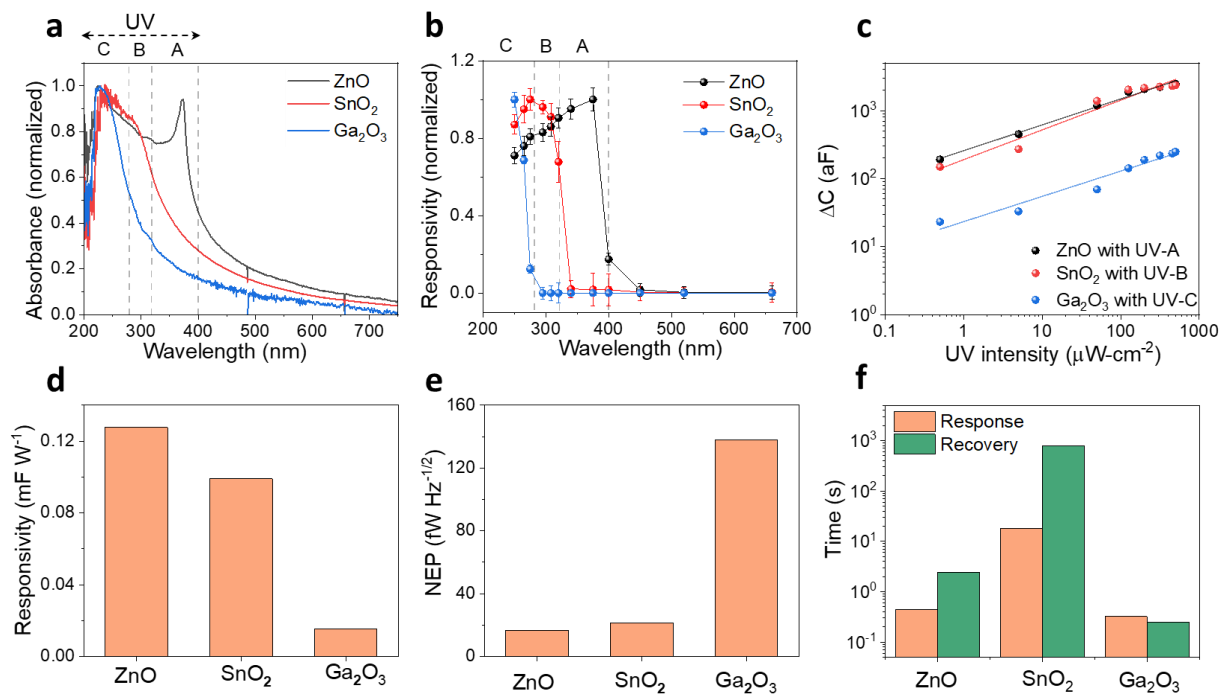

**Figure 2**. Evaluation of UV-detection performance of the capacitive sensor arrays. a) Normalized absorbance spectra of ZnO, SnO$_2$ and Ga$_2$O$_3$ NP dispersions in water. b) Normalized responsivity of the NPs at different exposure wavelengths. Regions corresponding to different UV bands (A, B, and C) are indicated by dashed lines in both a and b. c) Capacitance-changes in ZnO, SnO$_2$ and Ga$_2$O$_3$ NPs under varying UV-A, -B, and -C intensities, respectively. d) Optimal values of responsivity and e) noise-equivalent power (NEP) achieved with the ZnO, SnO$_2$ and Ga$_2$O$_3$ NPs under UV light (A, B, C, respectively). f) Response and recovery times of the ZnO, SnO$_2$ and Ga$_2$O$_3$ NP-coated pixels. ZnO and SnO$_2$ chips were cured at 100 °C and 200 °C, respectively for 10 minutes. Ga$_2$O$_3$ chip was cured at room-temperature.

With these spectral windows established, we quantified the dependence of sensor output on incident UV intensity; a critical parameter for accurate dosimetry and imaging. ZnO, SnO$_2$ and Ga$_2$O$_3$ chips were illuminated with UV-A, UV-B, and UV-C, respectively at intensities from 0.5 to 500 µW cm$^{-2}$, and the most responsive pixels were identified (Figure 2c). ZnO and SnO$_2$ exhibited capacitance increases up to 13× (191 to 2462 aF) and 16× (149 to 2362 aF), respectively, while Ga$_2$O$_3$ increased ~11× under UV-C (23 to 248 aF), though, the absolute magnitudes were roughly one order lower than those of ZnO and SnO$_2$ (Figure 2c). These observed increments surpass the system's minimum quantifiable capacitance (3.3 aF) by ~7-746×, thereby ensuring reliable detection of low-intensity UV signals in the present CMOS sensor. We also observed that Matrix 1 consistently produced the strongest signals (Figure S13, Supporting Information) due to its larger channel volume and was thus selected for all subsequent analyses.

From these optimized pixels, we determined maximum responsivities of 0.13, 0.10, and 0.015 mF W$^{-1}$ for ZnO, SnO$_2$, and Ga$_2$O$_3$, respectively (Figure 2d, Figures S14, S15, Supporting Information). It is important to note that the responsivity unit in capacitive sensors (F W$^{-1}$) is different from that used for conventional charge-based photodetectors (A W$^{-1}$; Supplementary Table T1, Supporting Information), and, therefore it is better to compare sensor performance by their noise-equivalent power (NEP). The corresponding NEP values were; 17, 22, and 138 fW Hz$^{-1/2}$ (Figure 2e, Figure S16, Supporting Information), comparable to those of state-of-the-art UV detectors that typically range from a few to several thousand fW Hz$^{-1/2}$ (Supplementary Table T1, Supporting Information). Finally, we evaluated the temporal response to assess suitability for real-time imaging. Ga$_2$O$_3$ pixels exhibited fast and highly repeatable dynamics, with response and recovery times of 0.32 s and 0.25 s, respectively (Figure 2f, Figure S17, Supporting Information), limited primarily by the microcontroller sampling rate. Room-temperature cured ZnO and SnO$_2$ films were slower (Figures S18a, S19a, Supporting Information) due to trap-mediated charge dynamics; that were subsequently improved by thermal curing. ZnO pixels cured at 100 °C demonstrated markedly improved dynamics owing to defect-state removal,[46] with response and recovery times of 0.44 and 2.45 s, respectively; 1-2 orders of magnitude faster than the as-deposited device (Figure S18, Supporting Information). SnO$_2$ required 200 °C curing to achieve its fastest response (~18 s, Figure S19, Supporting Information), although residual deep-trap effects remained, likely due to defects in the NPs and might be improved with further research.[47]

To understand how the effective relative permittivity ($\varepsilon_{\text{eff}}$) of the deposited MOX NPs influences the capacitance of the sensor pixels, we performed finite-element simulations using COMSOL Multiphysics (Supplementary Note 3 and Figure S20, Supporting Information). The simulated trends aligned well with the results shown in Figure 2c. These insights allow the capacitance change to be expressed as $\Delta C \propto \varepsilon_{\text{eff}}(I)$, where $\varepsilon_{\text{eff}}$ varies with UV intensity ($I$).

Collectively, these results confirm the origin of observed UV-response from the PDE, which is strongly correlated with the bandgap of the sensing material. Moreover, insights from these drop-cast experiments motivated us to explore more precise and scalable functionalization strategies to enable UV imaging with the capacitive CMOS arrays.

## 2.3. Multispectral imaging with inkjet-printed sensors

Multispectral imaging on CMOS platforms requires precise, scalable functionalization of individual pixels with wavelength-selective materials, which cannot be reliably achieved through drop casting. To meet this requirement, we adopted inkjet printing as an alternate strategy for spatially selective deposition (**Figure 3**). Beyond its sustainability and material efficiency, inkjet printing enables precise placement of picolitre-scale droplets,[48] offering a route to integrate ultralow quantities of functional materials directly onto CMOS pixels. Although inkjet patterning of CMOS substrates has been demonstrated for humidity sensing using graphene and metal-oxide systems,[49,50] its potential for CMOS-integrated photodetection has remained largely unexamined.

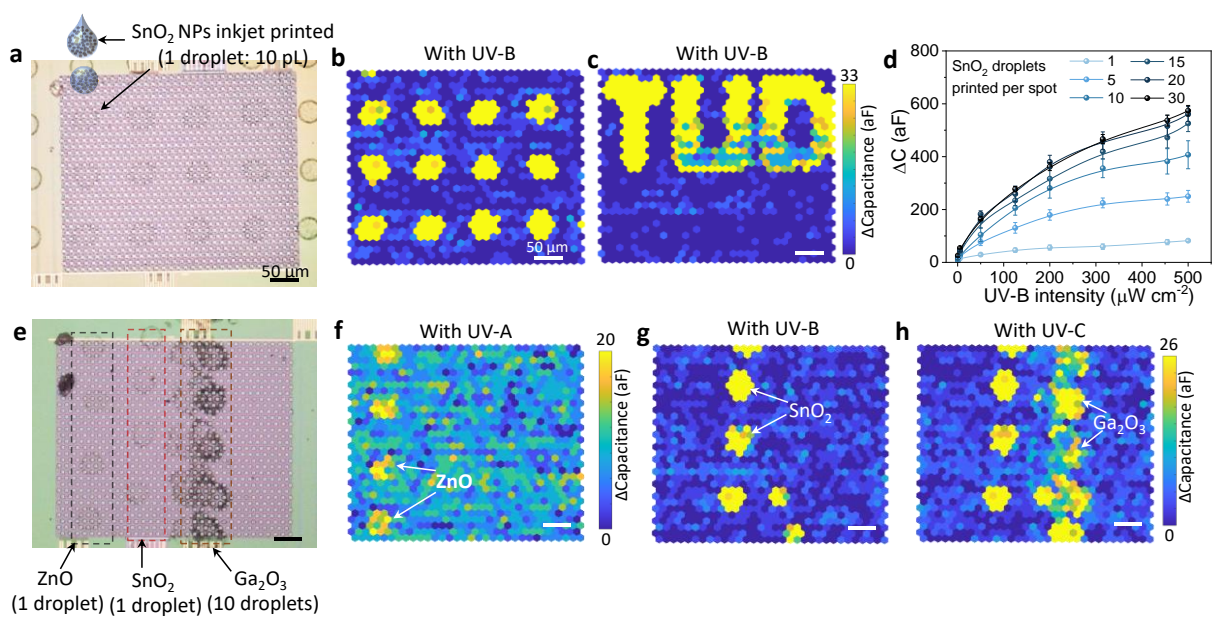

**Figure 3**. Capacitive CMOS sensor functionalized by inkjet printing. a) Optical microscopy image of Matrix 1 inkjet-coated with $SnO_2$ NP droplets (1 droplet with a volume of 10 pL) in a controlled grid. A schematic representation of inkjet droplets is shown above the figure. b) Response of Matrix 1 exposed under UV-B. Only $SnO_2$-functionalized pixels show response under UV-B. c) Response with UV-B, where the letters "TUD" were inkjet printed on Matrix 1 using the $SnO_2$ ink droplets. d) Capacitance variation of the sensor pixels inkjet-coated with different numbers of $SnO_2$ ink droplets (1 - 30) per spot under varying UV-B intensities (0.5 – 500 µW cm$^{-2}$). e) Optical microscopy image of Matrix 1 inkjet-coated with 10 pL droplets of ZnO (1 droplet per spot), $SnO_2$ (1 droplet per spot) and $Ga_2O_3$ (10 droplets per spot), facilitating multispectral UV detection within a single array. Response of Matrix 1 with f) UV-A, g) UV-B and h) UV-C, respectively. The scale bars in c, e, f, g, and h represent 50 µm. The capacitance scale bar for b and c is shown to the right of c, and for g and h, to the right of h.

We therefore sought to exploit the synergy between the localization afforded by inkjet deposition and the intrinsic versatility of PCS-based photodetection. By doing so, we aimed to establish a generalizable method for creating highly resolved, on-chip patterns of UV-responsive materials that could enable multispectral imaging. To this end, we formulated a suite of inkjet-printable MOX inks (Figures S21, S22, S23, Supporting Information) and used them to deposit 10 pL $SnO_2$ droplets in a controlled grid on Matrix 1 (Figure 3a). This configuration allowed us to evaluate both the spatial precision of the printing process and the photodetection capabilities of individual, selectively coated pixels. Each droplet reliably covered 14-15 pixels, and an array of 16 droplets with ~100 µm spacing provided a well-defined testbed.

Under UV-B illumination (250 µW cm$^{-2}$), pixels coated with $SnO_2$ exhibited clear capacitive responses reaching 50 aF (Figure 3b). These measurements confirm that even picolitre-scale depositions; nearly five orders of magnitude smaller than conventional drop-cast volumes, contain sufficient NP loading to activate UV responsiveness at the pixel level. The positional accuracy of inkjet printing further enabled deliberate patterning: printing the letters "TUD" (Figure S24, Supporting Information) produced a latent image that became visible only under UV-B excitation (Figure 3c), illustrating the capacity to encode and reveal information on demand.

To understand the relationship between droplet morphology and device response, we examined the NP distribution within the printed droplets. SEM imaging revealed peripheral densification consistent with the coffee-ring effect (Figures S25, S26, S27, Supporting Information), which reduces uniformity when only a single droplet is deposited. By printing multiple droplets at the same location, we increased the NP density and improved the coverage uniformity, resulting in progressively larger photoresponses. For $SnO_2$, increasing the droplet count from 1 to 30 enhanced the UV-B-induced (500 µW cm$^{-2}$) capacitance change from 82 to 573 aF (Figure 3d, Figure S25, Supporting Information). This tunability demonstrates that inkjet printing not only offers spatial selectivity but also provides a means of controlling responsivity via local material loading.

Comparative experiments with ZnO and $Ga_2O_3$ inks highlight the generality of the approach while revealing material-specific performance limits. ZnO produced lower UV-A responses (13–71 aF, Figure S28a, Supporting Information), aligned with its reduced NP concentration (Figure S26, Supporting Information), whereas $Ga_2O_3$ droplets generated the smallest changes under UV-C (14–27 aF, Figure S28b, Supporting Information). Responsivity and NEP values followed the same trend (Figure S29,

Supporting Information) and, as expected, remained 2–3 orders of magnitude lower than those of drop-cast films due to the much lower NP loadings.

Building on these insights, we extended our approach to achieve multispectral UV detection within a single array. We deposited 10 pL droplets of ZnO (1 droplet), $SnO_2$ (1 droplet), and $Ga_2O_3$ (10 droplets) at predefined locations on Matrix 1 (Figure 3e), creating spatially isolated pixels with distinct spectral sensitivities. A greater number of $Ga_2O_3$ droplets was required to achieve measurable response due to its intrinsically lower UV-C responsivity. This heterogeneous functionalization enabled selective detection of UV-A, UV-B, and UV-C illumination across the same device (Figure 3f–h). Together, these results validate inkjet printing as a practical and highly controllable route to pixel-resolved functionalization of CMOS imagers, providing the foundation for scalable multispectral UV sensing and enabling new modalities such as on-chip optical encoding.

### 2.4. Imaging with spin-coated sensors

Finally, we investigated the imaging capabilities of uniformly coated PCS arrays. Uniformity in pixel response is essential for accurate UV imaging, as variations in the sensing layer can directly translate into spatial artifacts in the reconstructed images. To ensure consistent pixel behavior, we implemented an easily applicable spin-coating approach to deposit MOX NP films across the PCS arrays (**Figure 4**). This method produced highly uniform $SnO_2$ coatings (Figure 4a), which in turn enabled consistent UV-B responses across all three matrices (Figure 4b and Figure S30, Supporting Information). Matrix 1, benefiting from its wider channels, facilitates more even NP spreading, and delivers the highest uniformity, with pixel responses clustered tightly around $33 \pm 5$ aF (Figure 4c,d). A similarly uniform response was obtained using spin-coated ZnO films, with UV-A sensitivity reaching $58 \pm 10$ aF (Figure 4d). $Ga_2O_3$, which exhibits intrinsically weaker UV responsivity, were not compatible with spin coating as it required a thicker film to generate measurable signals. For this reason, we employed drop-casting to achieve higher material loading (Figure S10, Supporting Information). While this approach yielded sufficient sensitivity under UV-C illumination, the resulting films were less homogeneous, leading to a broader response distribution ($162 \pm 32$ aF, Figure 4d).

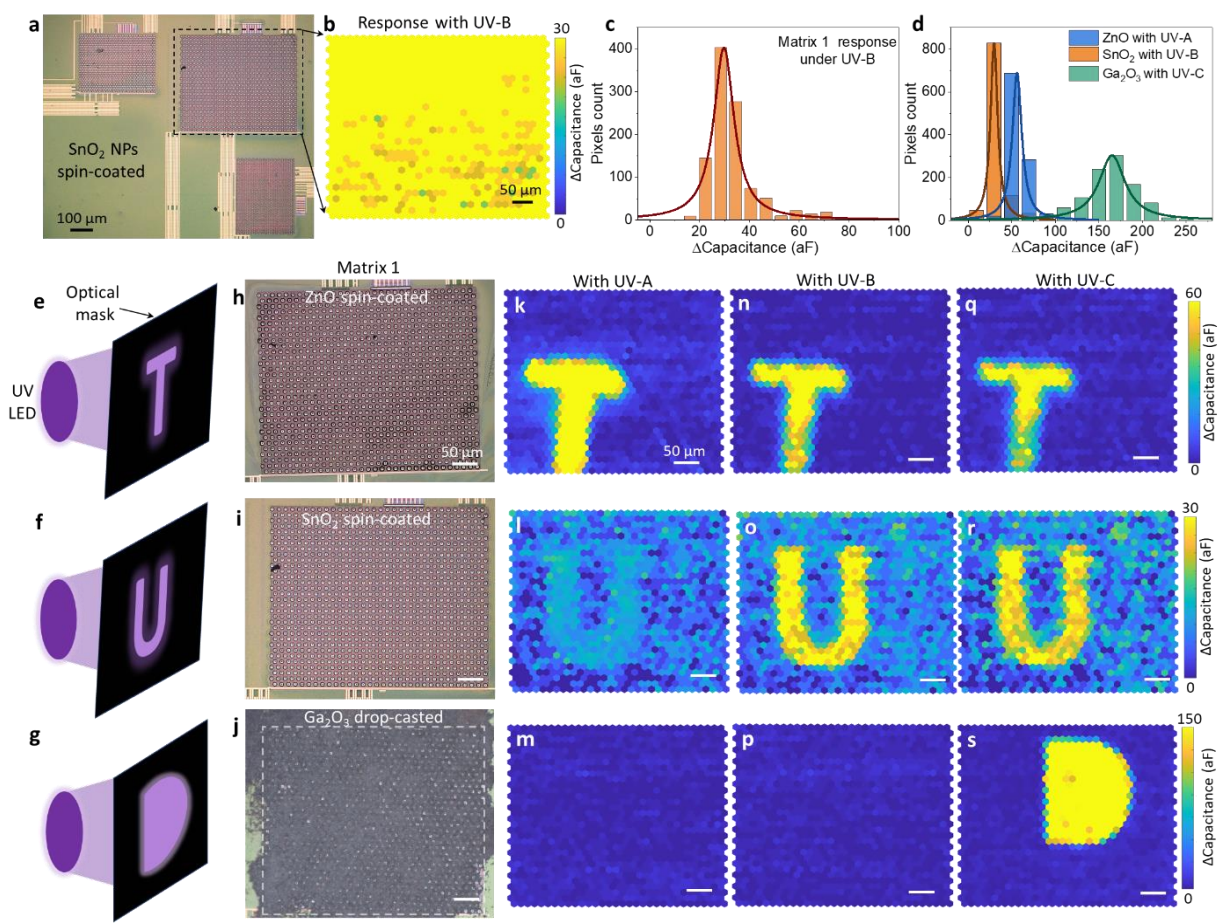

**Figure 4.** Capacitive CMOS sensor functionalized by spin coating. a) Optical microscopy image of $SnO_2$-NPs spin-coated on PCS arrays. b) Capacitive response of $SnO_2$-coated Matrix 1 under UV-B exposure (0.25 mW cm$^{-2}$), indicating uniform responses from the majority of the pixels. c) The distribution of capacitive responses in Matrix 1 of the $SnO_2$ spin-coated chip under UV-B light. d) The distribution of pixel responses for ZnO, $SnO_2$ spin-coated and $Ga_2O_3$ drop-casted chips under UV-A, UV-B and UV-C illumination, respectively, at similar intensities (0.25 mW cm$^{-2}$). e-g) Schematics of the optical shadow masks of the letters "T", "U" and "D" for UV light to pass through and projected on Matrix 1 of h) ZnO, i) $SnO_2$ spin-coated and j) $Ga_2O_3$ drop-coated chips. The response of ZnO, $SnO_2$ and $Ga_2O_3$ coated pixels under k-m) UV-A, n-p) UV-B and q-s) UV-C, respectively. Scale bars in i, j, l-s represent 50 μm. The capacitance scale bar for k, n and q is shown to the right of q; for l, o and r is shown to the right of r; and for m, p and s is shown to the right of s.

With these optimized sensing layers, we next evaluated the ability of the PCS arrays to resolve spatial UV patterns (Figure 4e-s). Each film was illuminated through a laser-milled metallic shadow mask containing the letters "T", "U", and "D" (Figure S31, Supporting Information). This enables direct assessment of spectral selectivity and imaging fidelity. ZnO which is responsive across a broad UV spectrum, clearly resolved the letter "T", most prominently under UV-A. $SnO_2$, which is selectively sensitive to UV-B, produced a strong and well-defined "U", while $Ga_2O_3$ responded exclusively under UV-C, enabling sharp visualization of the letter "D". The slightly enlarged appearance of the reconstructed features is likely due to variations in the angles of incidence of the light-beams that pass through the metal masks.

Furthermore, to probe the dynamic performance, we tracked a moving UV source across the chip surface (Supplementary Note 4, Figures S32, S33, Supporting Information). The device successfully captured temporal shifts in illumination, endorsing the capability of the PCS platform for real-time UV imaging, in which the sensing modality can be tuned simply by selecting the appropriate oxide film.

## 3. Conclusion

Here, we developed capacitive CMOS sensors for multispectral UV imaging that contain 32×32 pixels each in three individually addressable arrays, and allow use-case-specific late-functionalization of the chip surface outside the foundry. To activate the sensors for selectively detecting different UV bands, we utilized wide-bandgap semiconducting MOX NPs. Unlike conventional charge-based imagers, the present device employs a capacitive operation principle where UV light causes capacitance changes in the functionalization layer via PDE. Under UV exposure (500 µW cm$^{-2}$), MOX-functionalized pixels produce high capacitance variations (up to 2462 aF), which can be readily detected, as these responses are nearly 746× the lowest quantifiable capacitance-response. Across UV bands, the sensors exhibit low NEP (17-138 fW Hz$^{-1/2}$), comparable to state-of-the-art UV detectors, which typically range from a few to several thousand fW Hz$^{-1/2}$. Furthermore, the use of precision inkjet printing enables highly localized pixel-level functionalization and supports multispectral UV detection within a single array, demonstrating the versatility of the capacitive-CMOS platform even with ultralow material volumes (10 pL). Additionally, spin-coating of functional MOX layers enables uniform, high-resolution, and band-selective UV imaging without relying on intricate high-temperature processes typically required for standard CMOS imagers. This work, thus, establishes a scalable, sustainable, and fully CMOS-compatible strategy for fabricating low-cost multispectral UV imagers with the potential for extension to other spectral regions. Beyond imaging, the present work also opens new avenues for developing versatile, monolithic multisensing platforms, where integrating diverse functional materials could enable detection of a wide range of external stimuli, paving the way for future applications.

## 4. Experimental Section

*Deposition of metal-oxide nanoparticle dispersions*

Dispersions (5 wt%) of ZnO, SnO$_2$ (~100 nm), and Ga$_2$O$_3$ nanoparticles (Sigma-Aldrich) were prepared in deionized water and ultrasonicated for 2 hours. The dispersions of volume ~1 µL were drop-cast onto the chips. To deposit 3 different types of MOX NPs onto a single chip, we used syringes mounted on a Hyrel printer equipped with a micropositioner and optical camera, which allows accurate dispensing (~1 µL) onto defined regions of the PCS arrays.

*Spin coating of metal-oxide nanoparticle inks*

ZnO (in isopropyl alcohol) and SnO$_2$ (in butanol) inks (particle size <20 nm, Sigma-Aldrich) were used without modification. Uniform films were formed by spin-coating 1 µL of the ink at 1500 rpm for 1 min, followed by room-temperature drying for 1 hour.

*Inkjet printing of metal-oxide nanoparticle inks*

Inkjet printable ink-solutions were prepared by dispersing ZnO, SnO$_2$, and Ga$_2$O$_3$ NPs (5 wt%) in a 1:9 ethanol-water mixture. To lower the surface tension of the ink, sodium dodecyl sulfate (0.2 wt%) was added as a surfactant. The mixture was then ultrasonicated for 4 hours and subsequently, stirred for 4 days to yield stable NP ink-solutions. Dynamic light scattering (DLS) confirmed the presence of well-dispersed ZnO, SnO$_2$, and Ga$_2$O$_3$ nanoparticles with average hydrodynamic particle diameters of 314, 338, and 325 nm, respectively (Figure S21, Supporting Information), which are well below the 22 µm nozzle-width of the inkjet printer, minimizing the risk of clogging[48]. The viscosity (1.05-1.1 mPa s, Figure S22, Supporting Information) and surface tension (33-38 mN m$^{-1}$, Figure S22, Supporting Information) of the prepared inks also fall within the printable range[48], enabling the formation of stable, well-defined and repeatable inkjet droplets (Figure S23, Supporting Information). Inkjet printing was performed using a PiXDRO LP50 inkjet printer with DMC-11610 cartridges which produced 10 pL droplets.

*Photodetection measurements*

Photodetection was carried out using LEDs (Thorlabs) with defined peak wavelengths. For the experiments (except Figure 2b), we employed LEDs with peak wavelengths of 375 nm, 308 nm, and 255 nm to represent UV-A, UV-B, and UV-C radiation, respectively. LED-intensities were tuned with a neutral density filter and calibrated using a Thorlabs PM100A power meter coupled to an S401C sensor. The PCS arrays were then exposed to the LEDs, and capacitance values were recorded with measurement board (Figure S6, Supporting Information). Prior to the experiments, packaged PCS chips (received from NXP Semiconductors) were cleaned with deionized water and isopropyl alcohol.

*Calculation of Responsivity and Noise-equivalent Power (NEP)*

The responsivity ($R$) in F W$^{-1}$ was calculated using the following formula:

$$R = \frac{\Delta C}{P \times A}$$

where, $\Delta C$ is the capacitance change under UV exposure with respect to the capacitance under dark conditions, P is the intensity of the incident light, and A is the active area under illumination ≈ 300 µm$^2$ (see Figure S16, Supporting Information).

The NEP in W Hz$^{-1/2}$ was calculated using the following formula:

$$NEP = \sigma_{noise} \times \sqrt{\frac{2}{f_{sampling}}} \times \frac{1}{R}$$

where, $f_{sampling}$ is the sampling rate = 4.55 Hz, $\sigma_{noise}$ is the standard deviation of measured noise = 3.21 aF (Figure S16, Supporting Information).

**Supporting Information**

Supporting Information is available from the Wiley Online Library or from the author.

**Acknowledgement**

Authors acknowledge the facilities of the Department of Microelectronics (Bioelectronics Section) and the Department of Precision and Microsystems Engineering, TU Delft. S.K. and F.P. W. acknowledge funding from the European Union Horizon 2020 programme for project Distributed Artificial Intelligent System (DAIS), Grant agreement number: 101007273. P.G.S. and M.K.G. acknowledge support from the Dutch government as part of the National Growth Fund programme NXTGEN Hightech, project Agrifood 11: Greenhouse Horticulture - Digital Twin.

**Conflict of Interest**

The Authors declare no conflict of interest.

**Data Availability Statement**

The data that support the findings of this study are available from the corresponding authors upon reasonable request.